\documentclass[twocolumn,showpacs,preprintnumbers,amsmath,amssymb]{revtex4}

\usepackage{graphicx}
\usepackage{dcolumn}
\usepackage{bm}

\begin{document}

\preprint{APS/123-QED}

\title{Energy non-equipartition in multicomponent granular mixtures}

\author{Renaud Lambiotte}
\email{Renaud.Lambiotte@ulg.ac.be}

\author{L\'eon Brenig}
\email{lbrenig@ulb.ac.be}

\affiliation{ Physique Statistique, Plasmas et Optique Non-lin\'eaire, Universit\'e Libre de Bruxelles, Campus Plaine, Boulevard du Triomphe, Code Postal 231, 1050 Bruxelles, Belgium \\
}

\date{09/07/2005}

\begin{abstract}
We study non-equipartition of energy in granular fluids composed by an arbitrarily large  number of components. We focus on a simple mean field model, based upon a Maxwell collision operator kernel, and predict the temperature ratios for the species. Moreover, we perform  Direct Monte Carlo simulations in order to verify the predictions.
\end{abstract}


\maketitle

\section{Introduction}

The equipartition principle states that the total energy of an equilibrium macroscopic system is equally distributed among the degrees of freedom of its components.  For instance, in a classical gas mixture, equilibrium statistical mechanics show that equipartition of energy between the components is overwhelmingly more probable than a state characterized by different energies, i.e. the entropy of the system reaches its maximum at the equipartition state \cite{kac}. This result ensures the  occurrence of energy equipartition at the thermodynamic limit, and allows to deduce a microscopic expression and interpretation for the thermodynamic temperature. In this context, temperature is found to be the average kinetic energy of a molecule per degree of freedom. This property, which leads to the kinetic definition of temperature, reads $T_E = \frac{1}{Nd}<K>_{E}
$, where $Nd$ is the number of degrees of freedom in the system, and $<K>_E$ is an equilibrium average of the kinetic energy. 
This preliminary discussion shows the central role played by equipartition of energy in the foundations of equilibrium statistical mechanics, and its deep links with fundamental equilibrium concepts such as energy bath or thermal equilibrium. At the macroscopic level,  equipartition of energy is equivalent to the {\em zeroth law} of thermodynamics, which  states that if two systems are  in thermal equilibrium with a third system, they are in thermal equilibrium with each other.  

However, the above kinetic definition  may also be applied  in non-equilibrium situations, $
T_{NE} = \frac{1}{Nd}<K>_{NE}
$.
This kinetic temperature, which is now defined as an average with a non-equilibrium distribution in phase space, is a  less fundamental quantity than the equilibrium temperature, and is equivalent  to the average agitation of the particles in the system. By construction, it does not highlight a priori equilibrium features like the equipartition principle.   Nonetheless, its inclusion in systems
  where thermodynamics  are ill-defined  may be  relevant practically, as this quantity still provides meaningful information about the variance of the velocity distribution and about the local average energy in the system.  This approach has been applied during the last years to describe granular fluids, namely systems composed by a large number of agitated grains  interacting inelastically.
In this context,  
   the {\em granular temperature} is defined kinetically by:
 \begin{equation}
 \label{kineticDefinition}
   T({\bf r}, t) = <\frac{1}{d}\frac{m  V^{2}}{2}>
   \end{equation}
   where ${\bf V}$=${\bf v} - {\bf u}({\bf r}, t) $ is the random velocity, {\bf u}({\bf r}, t) is the local mean velocity and d is the dimension of the system. The average (\ref{kineticDefinition}) is now performed with the one particle velocity distribution $f({\bf r}, {\bf v}; t)$. This quantity has been shown to give a relevant  description of granular fluids, for instance by paving the way to an hydrodynamic formalism. 
In contrast,  however, the granular temperature exhibits non-equilibrium features, such as an anomalous Fourier law for the energy flux \cite{soto1}, and non-equipartition of energy in granular mixtures. This non-equipartition phenomenon has been first  theoretically predicted in the case of binary mixtures by \cite{garzo5}, and successfully verified by numerical simulations \cite{dahl} and experiments \cite{feitosa1}.
 It has also been observed in the case of rough inelastic spheres, where  the {\em translational temperature} and the {\em rotational temperature} are different. It is important to note here that energy non-equipartition deserves a careful study due to  macroscopic consequences such as non-negligible corrections to the transport coefficients \cite{garzo2}, and its role in vertical segregation \cite{dahl0}.

In this paper, we introduce a coherent mean field model for low density granular fluids composed by an arbitrary large number of components. This model rests on a  detailed analysis for the collision frequencies of the different components, and generalizes Maxwell-like models introduced by Ben-Naim \cite{ben3} and Marconi \cite{mar2}.
The values for energy non-equipartition are derived analytically in the low inelasticity limit, and are successfully compared with inelastic hard spheres theory and computer simulations. 
   
 \section{Two Rates Maxwell Model}  
In this paper, we consider a binary system composed by smooth inelastic hard spheres in $d$ dimensions, i.e. the interactions are instantaneous when the spheres are in contact and there is no transfer of angular momentum. Moreover, we focus on a system that is and remain spatially homogeneous in order to focus on the Homogeneous Cooling State (HCS). There are K kinds of grains, which are characterized by the following mechanical quantities: their respective mass $m_i$ and diameter $\sigma_i$ as well as their mutual inelasticity coefficients $\alpha_{ij}$ that describes the energy dissipation during a collision between particles i and j. These different properties  discriminate the components, which are described at the macroscopic level by their concentration $x_i = \frac{N_i}{N}$ and partial temperature $T_i \equiv  <\frac{1}{d}\frac{m_i  V^{2}}{2}>_i$, where the average is performed over species $i$. By definition, these macroscopic quantities are constrained by the relations:
$\sum_{i=1}^K x_i = 1$ and $\sum_{i=1}^K x_i T_i=T$.
Conservation of impulsion implies the following collision rule:

\begin{eqnarray}
\label{collisionRule}
{\bf v}_{i}^{*} &=& {\bf v}_{i} -  \frac{m_{j}}{(m_{i}+m_{j})}  (1+ \alpha_{ij} ){\mbox{\boldmath$\epsilon$}}
({\mbox{\boldmath$\epsilon$}}.{\bf v}_{ij}) \cr
{\bf v}_{j}^{*} &=& {\bf v}_{j} + \frac{m_{i}}{(m_{i}+m_{j})}  (1+ \alpha_{ij} ){\mbox{\boldmath$\epsilon$}}
({\mbox{\boldmath$\epsilon$}}.{\bf v}_{ij}) 
\end{eqnarray} 
where ${\bf v}_{ij}$ and ${\bf r}_{ij} $ are respectively the relative velocities ${\bf v}_{ij}$=${\bf v}_{i} - {\bf v}_{j}$ and positions ${\bf r}_{ij}$=${\bf r}_{i} - {\bf r}_{j}$ of the colliding discs i and j.
The star velocities are their post-collisional velocities and
${\mbox{\boldmath$\epsilon$}}$ is the unitary vector along the axis joining the centers of the two colliding
spheres.
In the low density limit, by assuming that pre-collisional correlations may be neglected \cite{soto}, the system is 
described by a system of $K$ coupled Boltzmann equations.
In this paper, we use mean field methods in order to simplify the mathematical structure of the collision operator, by
assuming that the collision frequency between particles i and j, i.e. proportional to $v_{ij}$ in the case of hard spheres, is approximated by the mean field quantity $\overline{\nu_{ij}} $:
\begin{equation}
\label{assumption}
\overline{\nu_{ij}} = \frac{1}{\sqrt{2} }\sqrt{\frac{T_i}{m_i} + \frac{T_j}{m_j}}
\end{equation}
Let us stress that this kind of assumption is common in order to build Inelastic Maxwell Models \cite{balda},\cite{ben1}, \cite{mar2}. In the case of binary mixtures, a simplified form has been introduced by Ben-Naim and Krapivsky \cite{ben1}, $  \overline{\nu_{ij}}= \frac{1}{\sqrt{2} }\sqrt{T_i(t)+ T_j(t)}
 $.
 Nonetheless, this approximation does not describe correctly the mass dependence of the collision frequencies and does not give  correct relations in the case of mass-dispersed mixtures.
The assumption (\ref{assumption}) leads to a system of $K$ kinetic equations for the velocity distributions $f_i$:
\begin{eqnarray}
\label{mixtureAM2}
 \frac{\partial f_{i}({\bf v})  }{\partial t}  =  x_i \sigma_i^{d-1} \sqrt{\frac{T_i}{m_i}} ~ K_{ii}[f_{i},  f_{i}] \cr 
 + \sum_{j=1, j\neq i}^K \frac{x_j}{\sqrt{2}} \sigma_{ij}^{d-1} \sqrt{\frac{T_i}{m_i} + \frac{T_j}{m_j}} ~ K_{ij}[ f_{i},  f_{j}]
 \end{eqnarray} 
 where angular integrations have been absorbed into the time scale.
The integer i goes from 1 to K, and $K_{ij}$ is defined by:
\begin{equation}
\label{operator}
K_{ij}[f({\bf v}), g({\bf w})] = \int d{\mbox{\boldmath$\epsilon$}} d{\bf w} \frac{1}{\alpha_{ij}} [f({\bf v}^{'}) g({\bf w}^{'}) -f({\bf v}) g({\bf w})]
\end{equation}
Because of the Maxwell-like structure of the collision operators, this set equation leads to a closed system of equations for the partial temperatures $T_i$. Straightforward calculations lead to the following expressions:

 \begin{eqnarray}
  \label{tmixtureMulti}
\frac{\partial T_i}{\partial t}  = - x_i \sigma_i^{d-1} \sqrt{\frac{T_i}{m_i}} \frac{(1-\alpha_{ii}^2)}{2} T_i(t)
\cr + \sum_{j=1, j\neq i}^K \frac{x_j}{\sqrt{2} }\sigma_{ij}^{d-1} \sqrt{\frac{T_i}{m_i} + \frac{T_j}{m_j}}  \mu_{ji} (1+\alpha_{ij}) \cr \times [(\mu_{ji} (1+\alpha_{ij}) -2) T_i  +  \mu_{ij}  (1+\alpha_{ij}) T_j ]
 \end{eqnarray}
 where we have introduced the normalized mass ratios $\mu_{ij}=\frac{m_{i}}{m_i+m_j}$.
We rewrite this set into a more compact expression after defining related $\epsilon_{ij} = 1 - \alpha_{ij}$:
  \begin{eqnarray}
  \label{tmixtureN2}
\frac{\partial T_i}{\partial t}  =\sum_{j}^K x_j \frac{\sigma_{ij}^{d-1}}{\sqrt{2} }\sqrt{\frac{T_i}{m_i} + \frac{T_j}{m_j}}  \mu_{ij} (2 - \epsilon_{ij}) \cr \times [(- 2 \mu_{ij} - \mu_{ji} \epsilon_{ij}) T_i ~ + ~( 2 \mu_{ij} - \mu_{ij} \epsilon_{ij}) T_j ]
 \end{eqnarray}
 
The next step consists in introducing the quantities $R_i =\frac{T_i}{T_1}$, which measure the departure from energy equipartition. Obviously, the $K$ quantities $R_i$, $i=1...K$,  are equal to 1 when equipartition takes place, and $R_1$ is always equal to 1 by construction. We  rescale  the time d$\tau =$ dt$\sqrt{\frac{T_1(t)}{m_1}}$, thereby considering a time scale proportional to the average number of collision 1-1 in the system. Let us note that this particular choice is arbitrary. This leads to:

\begin{eqnarray}
  \label{tmixtureN4}
\partial_\tau R_i  
= \sum_{j}^K x_j \frac{\sigma_{ij}^{d-1}}{\sqrt{2} }\sqrt{R_i \frac{m_1}{m_i} + R_j \frac{m_1}{m_j}}  \mu_{ji} (2 - \epsilon_{ij}) \cr 
\times [(- 2 \mu_{ij} - \mu_{ji} \epsilon_{ij}) R_i ~ + ~( 2 \mu_{ij} - \mu_{ij} \epsilon_{ij}) R_j ] \cr
- R_i \sum_{j}^K x_j \frac{\sigma_{1j}^{d-1}}{\sqrt{2} }\sqrt{1 + R_j \frac{m_1}{m_j}}  \mu_{j1} (2 - \epsilon_{1j}) \cr 
\times [(- 2 \mu_{1j} - \mu_{j1} \epsilon_{1j})  ~ + ~( 2 \mu_{1j} - \mu_{1j} \epsilon_{1j}) R_j ]
 \end{eqnarray}
where one verifies  that $R_1 = 1$ as imposed by construction.
In the sequel, we are interested in the stationary solutions for small inelasticity parameters. Therefore, we solve (\ref{tmixtureN4}) by perturbation methods assuming that the quantities $\epsilon_{ij}$  are all small and of the same order of magnitude. The temperature ratios are written under the form  $R_j = 1 + \epsilon R_j^1$, where $\epsilon$ is a small formal perturbation coefficient. This development is based on the fact that the zero-th order solution of (\ref{tmixtureN4}) is $R_j=1$, $\forall j$, which corresponds to  equilibrium equipartition of energy. Given these quantities, the stationary first order solution of  (\ref{tmixtureN4}) reads:
\begin{eqnarray}
  \label{tmixtureN5}
 \sum_{j}^K x_j \frac{2 \sigma_{ij}^{d-1}}{\sqrt{2} }\sqrt{\frac{m_1}{m_i} +  \frac{m_1}{m_j}}  \mu_{ji} ~ ~   [ 2 \mu_{ij} (R_j^1 - R_i^1) -  \epsilon_{ij})  ] \cr
- \sum_{j}^K x_j \frac{2 \sigma_{1j}^{d-1}}{\sqrt{2} }\sqrt{1 +  \frac{m_1}{m_j}}  \mu_{j1} ~~  [ 2 \mu_{1j} R_j^1  - \epsilon_{1j})] = 0
 \end{eqnarray}
In a system composed by two components, calculations are straightforward and lead to the set of linear equations:
\begin{eqnarray}
\label{final}
  \frac{ 2 m_2 m_1}{(m_1+m_2)^2}   R^1_2 =  -  \frac{\sigma_{11}^{d-1}}{\sigma_{12}^{d-1}} \sqrt{\frac{m_1}{m_1+m_2}}\frac{x_2}{ \sqrt{2}} \epsilon_{22} \cr + \frac{\sigma_{22}^{d-1}}{\sigma_{12}^{d-1}}\sqrt{\frac{m_2}{m_1+m_2}} \frac{x_1}{ \sqrt{2}}  \epsilon_{11}
-    \frac{(m_1 x_1
-   m_2 x_2)}{(m_1+m_2)}    \epsilon_{12}  
 \end{eqnarray}
 where $R_1^1=0$ by definition.
This expression gives the first order deviation to equipartition, in the limit of small inelasticity.  Remarkably, solution (\ref{final}) is exactly equivalent to the solution obtained by Garzo and Dufty \cite{garzo5} from the  inelastic hard spheres Boltzmann equation in the low inelasticity limit. This equivalence, together with the mathematical simplicity induced by the mean field model, seem to indicate that the TRMM is a relevant model for granular mixtures, which could be used, for instance, in order to probe the applicability of hydrodynamics in these systems, or to derive  the temperature ratios in mixtures composed by an arbitrary number of components, as shown below. 

\section{Arbitrary number of components} 

In order to treat systems composed by a large number of species, it is helpful to rewrite (\ref{tmixtureN5}) into the canonical form: 
\begin{eqnarray}
  \label{tmixtureN8}
Ai R_i^1 +
 \sum_{j=2, j\neq i}^K B_{ij} R_j^1 = C_i
  \end{eqnarray}
where the coefficients read explicitly:
 \begin{eqnarray}
  \label{tmixtureN9}
A_i &=&  [2 x_i \sigma_{1i}^{d-1} \frac{ \sqrt{m_1 m_i} }{(m_1 + m_i)^{\frac{3}{2}}} \cr &+& \sum_{j=1, j\neq j}^K (2 x_j  \sigma_{ij}^{d-1} \frac{\sqrt{m_i m_j} }{(m_i + m_j)^{\frac{3}{2}}})] \cr
B_{ij} &=& -  2 x_j   [ \sigma_{ij}^{d-1}\frac{\sqrt{m_i m_j} }{(m_i + m_j)^{\frac{3}{2}}} - \sigma_{1j}^{d-1} \frac{\sqrt{m_1 m_j} }{(m_1 + m_j)^{\frac{3}{2}}}]  \cr
C_i &=& \sum_{j=1}^K x_j      ( \sigma_{1j}^{d-1} \sqrt{ \frac{m_j}{m_1}} \frac{1}{ \sqrt{m_1 + m_j}}  \epsilon_{1j} \cr &-& \sigma_{ij}^{d-1} \sqrt{\frac{m_j}{m_i}} \frac{1}{\sqrt{m_i+ m_j} }  \epsilon_{ij})  
 \end{eqnarray}
 Consequently, the whole dynamical problem is reduced to the inversion of the matrix {\bf M} defined by:
 \begin{equation}
M_{ij} = A_j \delta_{ij} + B_{ij} 
 \end{equation}
 Unfortunately, this problem is practically not trivial when the system is composed by a large number $K$ of components (inversion of a K-1 matrix) and leads to  analytical but intractable expressions for the $R_j$. In the following, we prefer to focus on a particular case which enormously simplifies the inversion of (\ref{tmixtureN8}) and whose solutions $R_j$ have a simple analytical form. Namely, we consider a system where all particles have the same mass $m_j=m$ and the same diameter $\sigma_{ij}=\sigma$, but whose inelasticities $\epsilon_{ij}$  vary.  
 Remarkably, in that case, all the non-diagonal terms $B_{ij}$ vanish exactly $B_{ij}=0$, so that the general solution of  (\ref{tmixtureN9}) reads:
  \begin{eqnarray}
  \label{multiFinal}
R_i^1 &=&   \sum_{j=1}^K x_j       (  \epsilon_{1j} - \epsilon_{ij})
 \end{eqnarray}
  This solution is instructive for several reasons. First, let us insist on the fact that it is the first expression for non-equipartition of energy in multi-component mixtures. This solution clearly shows that the  system behaves qualitatively in the same way as in a binary inelastic mixture, namely the partial temperatures $T_i$ do not reach asymptotically the same value, but that they remain proportional in the long time limit. This behaviour justifies  the derivation of granular hydrodynamics for complex mixtures. 
  Another nice feature of (\ref{multiFinal}) comes from its very simple expression, which allows to recover the solution (\ref{final}) when $K=2$, and which is easily generalized when the system is composed by an infinite number of species, namely when there is a continuum of inelasticities in the system. In that case, instead of the concentrations $x_i$ used in order to characterize the components, we use the concentration density $\rho(\iota)$, which represents the density of species $\iota$. The probability normalization reads:
   \begin{eqnarray}
  \label{tmixtureN12}
\int d\iota \rho(\iota)    =1
 \end{eqnarray}
Relation (\ref{multiFinal})  generalizes into:
  \begin{eqnarray}
  \label{tmixtureN13}
R(\iota) &=&   \int d\iota^{'} \rho(\iota^{'})       (  \epsilon_{r\iota^{'}} - \epsilon_{\iota \iota^{'}})
 \end{eqnarray}
 where r is the arbitrary chosen reference species to which the temperatures are compared, $R(\iota)=\frac{T(\iota)}{T(r)}$. 
 
 \begin{figure}
 
                \includegraphics[angle=-90,width=3.3in]{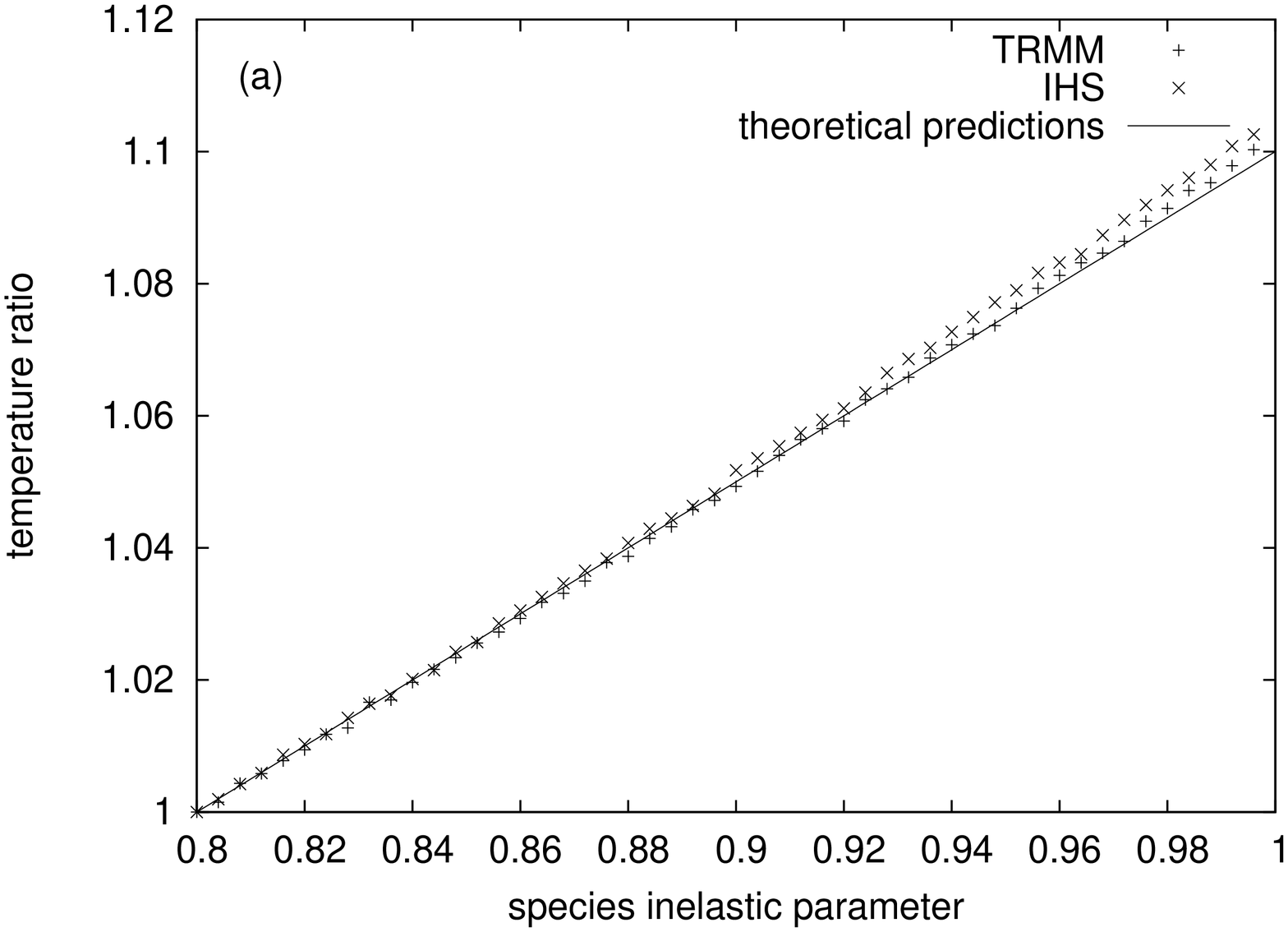}

              \includegraphics[angle=-90,width=3.3in]{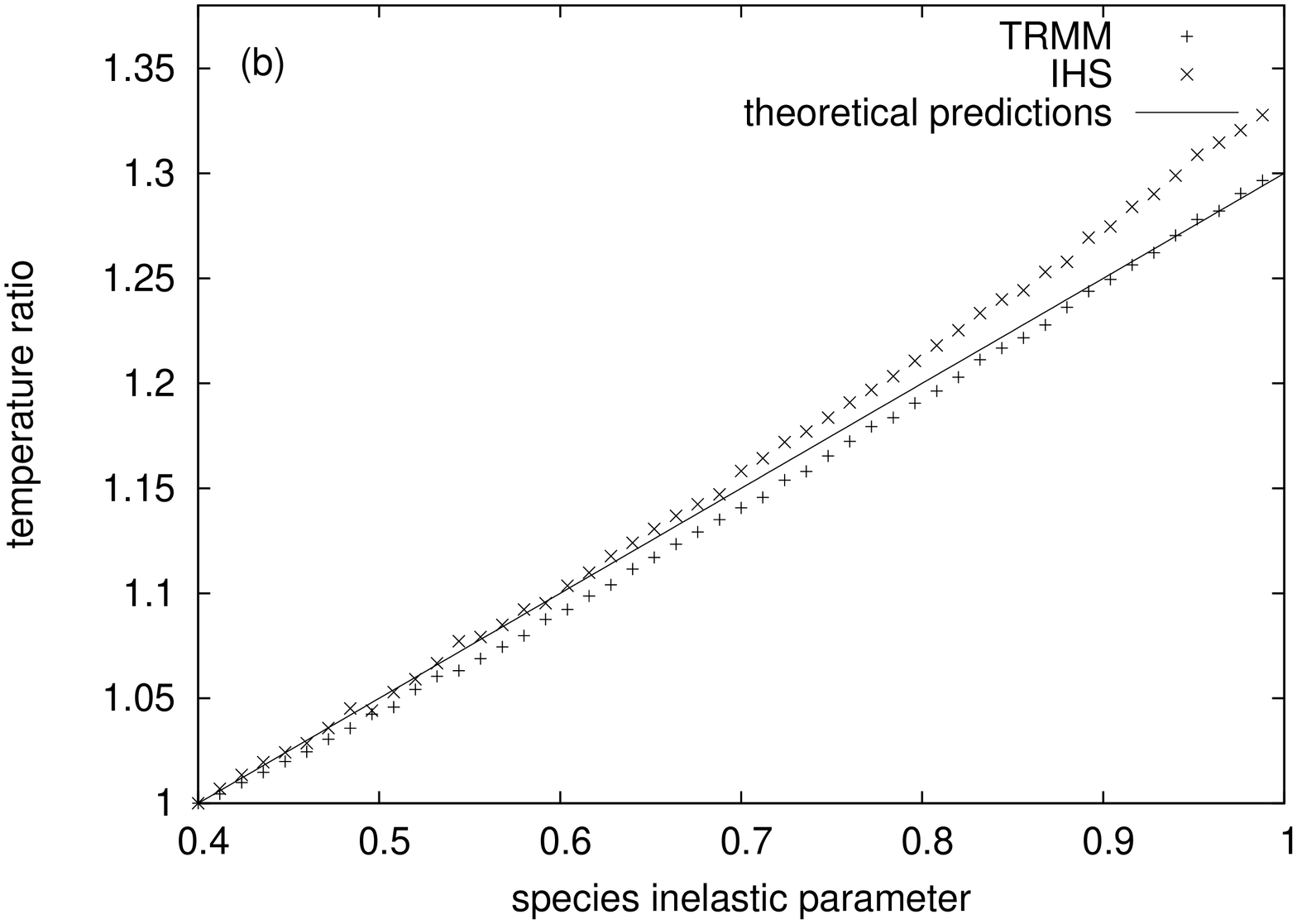}
                 
\caption{\label{figTRMM} \small{\em Asymptotic temperature ratio, as a function of the inelasticity of each species, in a system where $\alpha_{min} = 0.8$ (a) and $\alpha_{min} = 0.4$ (b). The solid line corresponds to the theoretical predictions (\ref{multiFinal2}), while the data points correspond to DSMC simulations of TRMM and IHS.}}
\end{figure}

We verify the validity of (\ref{multiFinal}) by performing two kinds of computer simulations for the multi-component mixture. The latter were performed for particles whose masses and diameters are equal, $m_i=1$ $\sigma_i=1$ in order to compare the results with (\ref{multiFinal}). First, we make DSMC simulations of the set of kinetic equations (\ref{mixtureAM2}), namely we apply the standard DSMC algorithm, where the probability of collisions of a pair is taken to be proportional to  $\sqrt{T_i + T_j}$. In the simulations, we considered system composed by a large number $K=50$ of components. Moreover, each species is composed by the same number of particles, $N_i=N_j$. This implies that the concentrations are equal to $x_i=\frac{1}{50}$. The species are discriminated by their inelasticities $\alpha_{ii}$, which are all assumed to be different, and we chose the reference component to be the one with the lower inelasticity: $\alpha_{11} < \alpha_{ii}$, $i> 1$. This arbitrary choice suggests that the quantities $R_i$ should be $\geq 1$ in the long time limit, given the fact that other components dissipate less energy than the reference component does. One should note, however, that the choice for the cross inelasticities $\alpha_{ij}$  can still be arbitrarily chosen provided they respect the symmetry relation $\alpha_{ij} = \alpha_{ji}$. In the following, we choose the following expression for the cross inelasticities:
 \begin{equation}
  \alpha_{ij} = \frac{\alpha_i + \alpha_j}{2} 
 \end{equation}
 which is a reasonable assumption for the inelasticity of collisions between particles i and j, and which allows to characterize each species i by their sole inelasticity $\alpha_{ii}$, instead of the vectorial quantity $(\alpha_{1i}, ..., \alpha_{ii}, ..., \alpha_{Ni})$. This assumption, which also implies that $ \epsilon_{ij} = \frac{\epsilon_i + \epsilon_j}{2} $, simplifies (\ref{multiFinal}) into:
 \begin{eqnarray}
  \label{multiFinal2}
R_j^1 &=&  \frac{1}{K} \sum_{n=1}^K       \frac{1}{2}(  \epsilon_1 + \epsilon_n - \epsilon_j - \epsilon_n)\cr
&=&  \frac{1}{2}(  \epsilon_1  - \epsilon_j )
 \end{eqnarray}
 In the simulation results presented below, we define a minimum and a maximum inelasticity in the system, $\alpha_{min}$ and $\alpha_{max}$ respectively. By definition, we chose $\alpha_{11}=\alpha_{min}$. Then, we fill uniformly the interval $[\alpha_{min}, \alpha_{max}[$ , i.e. $\alpha_{ii}=\alpha_{min} + (i-1) \delta \alpha$,
 where the quantity $\delta \alpha$ is defined by $\delta \alpha= \frac{(\alpha_{max}-\alpha_{in})}{50}$. Let us stress  that this uniform distribution  in the interval $[\alpha_{min}, \alpha_{max}[$ is an arbitrary choice, and that more general compositions can be considered without any analytical nor computational additional difficulty. In this work, we also perform DSMC simulations of the true set of inelastic Boltzmann equations for the same mixture, where no approximation is made to simplify the collision operator, thereby testing the validity of the TRMM.
 In figure \ref{figTRMM}, we present results in the small inelastic limit  $\alpha_{min} = 0.8$, that show an excellent agreement with (\ref{multiFinal2}), and in the high inelasticity limit $\alpha_{min} =0.4$, for which only small discrepancies from the predictions may occur, i.e. deviations do not exceed $15 \%$.

\section{Conclusion} 
In this paper, we have focused on the non-equipartition of energy in inelastic gases composed by a large number of species.  We have used mean field approximations, in order to simplify the set of Boltzmann equations, and to focus on an analytically tractable problem.  By doing so, we have defined the Two Rates Maxwell Model, that we have solved formally in the small inelasticity limit, thereby deriving the explicit values for energy non-equipartition in systems where components have the same mass and diameter.  
These predictions have been verified by simulations for systems composed by $50$ components, that show that the TRMM is a relevant model, even in very inelastic systems. At this point, it is important to note that the influence of the shape of the velocity distributions has not been taken into account by the mean field modeling. This is due to the Maxwell-like kernel of the collision operators, that leads to a closed set of equations for the partial temperatures, thereby neglecting influence of higher velocity moments \cite{lambi}.
  Nonetheless, this influence is usually weak and has no quantitative effect  when the system is weakly inelastic.
That was shown in the small inelasticity limit, both by the DSMC simulation results and by analytical comparisons  with the exact inelastic Boltzmann equation. 
As a consequence, the TRMM is an ideal candidate in order to apprehend more complex phenomena in  granular mixtures, such as hydrodynamics for a large number of components, or the detailed study of systems composed by a continuum species.

\acknowledgments
This work has been done thanks to a financial support of FRIA.
R.L. would like to thank warmly M. Mareschal for his invitation at CECAM, and  J. Wallenborn for stimulating discussions.

\end{document}